\documentclass[10pt,letterpaper]{article}
\usepackage{opex3}
\usepackage{color}
\usepackage{threeparttable}

\begin{document}

\title{Anisotropic diffraction induced by orbital angular momentum during propagations of optical beams }

\author{Guo Liang$^{1,2}$, Yuqing Wang$^{1,3}$, Qi Guo$^{1,*}$, and Huicong Zhang$^{1}$}

\address{$^1$ Guangdong Provincial Key Laboratory of Nanophotonic Functional Materials and Devices, South
China Normal University, Guangzhou 510631, P. R. China\\
$^2$School of Physics and Electrical Information, Shangqiu Normal University, Shangqiu 476000, China\\
$^3$School of Information Engineering, Wuyi University, Guangdong, Jiangmen 529020, P. R. China
}

\email{$^*$guoq@scnu.edu.cn} 



\begin{abstract}
It is demonstrated that the orbital angular momentum (OAM) carried by the elliptic beam without the phase-singularity can induce the anisotropic diffraction (AD).
 The quantitative relation between the OAM and its induced AD is analytically obtained by a comparison of two different kinds of (1+2)-dimensional beam propagations: the linear propagations
of the elliptic beam without the OAM in an anisotropic medium and that with the OAM in an isotropic one.
In the former case, the optical beam evolves as the fundamental mode of the eigenmodes when its ellipticity is 
the square root of the anisotropic parameter defined in the paper; while in the latter case, the fundamental mode exists
only when the OAM carried by the optical beam equals a specific one called a critical OAM. The OAM always enhances the beam-expanding in the major-axis direction and weakens that in the minor-axis direction no matter the sign of the OAM, and the larger the OAM, the stronger the
AD induced by it. Besides, the OAM can also make the elliptic beam rotate,
and the absolute value of the rotation angle is no larger than $\pi/2$ during the propagation.
\end{abstract}



\section{Introduction}\label{section1}
The diffraction phenomenon of light is described as the interference of electromagnetic waves according to the Huygens-Fresnel principle, which is a very old story back to the 15th century~\cite{Born-book}. Due to the diffraction, the (1+2)-dimensional paraxial optical beam expands its cross-sectional shape uniformly along the two transverse directions when it propagates in an isotropic bulk medium~\cite{Kogelnik-ao-1966,haus-book}. The situation 
in anisotropic media, either a uniaxial one~\cite{fleck-josa-83,Seshadri-josaa-2001,Ciattoni-josaa-2002,Ciattoni-pre-2002,Seshadri-josaa-2003,Ciattoni-oc-2004} or a biaxial one~\cite{Chen-oc-2011,Tinkelman-ol-2003}, however, is a little bit complicated.
This circumstance entails that the diffraction is anisotropic and does not exhibit the well-know circularly symmetric behavior typical of
the isotropic one~\cite{Ciattoni-oc-2004}. The beam loses its initial cylindrical symmetry during the propagation~\cite{Ciattoni-josaa-2002}, and
the fundamental (basic) mode among all eigenmodes~\cite{Kogelnik-ao-1966,haus-book} is the Gaussian beam with an elliptical cross section since only an elliptical
beam can preserve its cross-sectional shape on the propagation~\cite{Seshadri-josaa-2003}. There are two theoretical descriptions of the propagation in the anisotropic media. One of them is an integration of all of plane-wave
angular spectrums of the electrical field~\cite{Ciattoni-josaa-2002,Ciattoni-oc-2004, Ciattoni-pre-2002}, and the other is to solve the paraxial (differential) equation governing the beam propagation~\cite{fleck-josa-83,Seshadri-josaa-2001,Ciattoni-pre-2002,Chen-oc-2011} [see, Eq.~(\ref{equation-uniaxial-dimensionless}) in Sec.~\ref{section2}], in which there exists the anisotropic diffraction (AD)~\cite{Polyakov-pre-2002}, that is, the two coefficients (the Fresnel diffraction
coefficients~\cite{Chen-oc-2011}), $\delta_{xx}$ and $\delta_{yy}$, of the second derivative terms are different.

Different from a long history of diffraction, it was 1992 when Allen and coworkers~\cite{Allen-pra-1992} recognised that optical beams can carry a discrete orbital angular momentum (OAM). The concept of the OAM is now leading to new understanding of a
wide range of phenomena~\cite{Franke-Arnold-lpr-2008}, including fundamental processes in
Bose-Einstein condensates~\cite{Madison-prl-2000} and quantum information~\cite{MOLINA-TERRIZA-nphysics-2007}, while the associated technologies
have led to new applications in optical manipulation~\cite{Padgett-nphotonics-2011,Dholakia-NPhoton-2011}. The optical beams carrying the OAM are usually
associated with optical vortices and related ring-shaped beams with the phase-singularity, such as the Laguerre-Gauss beams~\cite{Allen-pra-1992} and the Bessel beams~\cite{Volke-Sepulveda-job-2002}, and so on. At the same time, the
vortex-free beam with nonzero OAM and without phase-singularity was also found~\cite{Courtial-oc-1997}, and is an elliptical Gaussian beam with the cross-phase term~\cite{Desyatnikov-prl-2010,liang-pra-2013}.
It has been shown that such elliptical Gaussian beams with nonzero OAM can survive in the form of the stably spiraling elliptic solitons in the nonlinear media with both linear and nonlinear isotropy~\cite{Desyatnikov-prl-2010, liang-pra-2013}.
 To explain the physical mechanism of such a phenomenon, we (two of the authors of this paper) guessed that~\cite{liang-pra-2013} the OAM could result in effective anisotropic diffraction, and went to explain qualitatively why the elliptic mode can survive in the medium with both linear and nonlinear isotropy where only the mode with cylindrical-symmetry is supposed to exist. This open question will be addressed in this paper.

The diffraction and the OAM are considered to be two essential and intrinsic properties of the many properties that optical beams possess.
Their interaction, that is the diffraction of the optical beam with the OAM, has been extensively exploited, and most of them focused on optical vortice diffraction in presence of obstacles, including a knife edge~\cite{Arlt-JMO-2003}, a single slit~\cite{Ferreira-OL-2011}, forked ~\cite{Saitoh-PRL-2013} and gradually-changing-period~\cite{Kunjian-OL-2015} gratings, triangular and square apertures~\cite{Hickmann-PRL-2010, Silva-OL-2014}, an annular aperture~\cite{Chengshan-OL-2009} and so on. But so far, there is no such an investigation, to our knowledge at least, that the optical beam with the OAM and without phase-singularity, for example the elliptical Gaussian beam with the cross-phase term, propagates in the linear bulk materials without obstacles.

In this paper, we discuss the linear propagations
of the elliptic beam without the OAM in the anisotropic bulk medium and that with the nonzero OAM in the isotropic one, respectively.
The AD induced by the OAM is analytically obtained by the comparison of two different kinds of propagations.
Besides, it is also found that the elliptic beam with the OAM will rotate during the propagation, and the rotation property is discussed in detail.

\section{Paraxial equation governing the beam propagation and its eigenmodes}\label{section2}
The paraxial optical beams propagating in the anisotropic bulk media is in general modeled by~\cite{Chen-oc-2011}
\begin{equation}\label{equation-anisotropic}
i\left(\frac{\partial \phi}{\partial \zeta}-\delta_\xi\frac{\partial \phi}{\partial \xi}-\delta_\eta\frac{\partial \phi}{\partial \eta}\right)+\frac{1}{2k_0}\left(\delta_{\xi\xi}\frac{\partial^2\phi}{\partial\xi^2}+\delta_{\xi\eta}\frac{\partial^2\phi}{\partial\xi\partial\eta}
+\delta_{\eta\eta}\frac{\partial^2\phi}{\partial\eta^2}\right)=0,
\end{equation}
where $\phi$ is the slowly varying amplitude for the paraxial beam, $\zeta$ is the longitudinal (the propagation direction of the beams) coordinate, $\xi$ and $\eta$ are the transverse coordinates normal to $\zeta$, $k_0=\left.k_\zeta(k_\xi,k_\eta)\right|_{k_\xi=k_\eta=0}$,
$\delta_i=\left.\partial_{k_i}k_\zeta(k_\xi,k_\eta)\right|_{k_\xi=k_\eta=0}$ and $\delta_{il}=\left.\partial^2_{k_ik_l}k_\zeta(k_\xi,k_\eta)\right|_{k_\xi=k_\eta=0}$ ($i,l$ means $\xi$ or $\eta$). The function $k_\zeta(k_\xi,k_\eta)$ can be derived from the dispersion equation to determine the plane wave modes with the form of $\exp[i(k_\zeta\zeta+k_\xi\xi+k_\eta\eta-\omega t)]$ existing in the anisotropic media~\cite{Chen-oc-2011,guo-joa-00}. $\delta_\xi$ and $\delta_\eta$ describe the walk-off of the beam center in $\xi$ and $\eta$ directions, and $\delta_{\xi\xi},\delta_{\xi\eta}$ and $\delta_{\eta\eta}$ are the Fresnel diffraction coefficients~\cite{Chen-oc-2011}. All of $\delta$-coefficients depend upon the direction of the propagation. For the uniaxial crystals, the paraxial equation for the extraordinary light is simplified to~\cite{fleck-josa-83,Chen-oc-2011,guo-joa-00}
\begin{equation}\label{equation-uniaxial}
i\left(\frac{\partial \phi}{\partial \zeta}-\delta\frac{\partial \phi}{\partial \xi}\right)+\frac{1}{2k_0}\left(\delta_{\xi\xi}\frac{\partial^2\phi}{\partial\xi^2}
+\delta_{\eta\eta}\frac{\partial^2\phi}{\partial\eta^2}\right)=0.
\end{equation}
In fact, the mixed derivative $\partial^2\phi/\partial\xi\partial\eta$ in Eq. (\ref{equation-anisotropic}) can be made vanished by the rotation of coordinates in the $\xi$-$\eta$ plane~\cite{Chen-oc-2011},
and then the paraxial equation for the optical beams in the  biaxial crystals [Eq.~(\ref{equation-anisotropic})] can also be reduced into the similiar form with Eq.~(\ref{equation-uniaxial}).
Using the transformation $x=(\delta\zeta+\xi)/{w_0},~y={\eta}/{w_0}$ and $z={\zeta}/{k_0w_0^2}$, the wave equation (\ref{equation-uniaxial}) becomes
its dimensionless form
\begin{equation}\label{equation-uniaxial-dimensionless}
i\frac{\partial \phi}{\partial z}+\frac{1}{2}\left(\delta_{xx}\frac{\partial^2\phi}{\partial x^2}
+\delta_{yy}\frac{\partial^2\phi}{\partial y^2}\right)=0,
\end{equation}
where $\delta_{xx}=\delta_{\xi\xi}$ and $\delta_{yy}=\delta_{\eta\eta}$. The last two terms of Eq.(\ref{equation-uniaxial-dimensionless}) are diffraction terms, and determine the expansion of the optical beams. That $\delta_{xx}\neq \delta_{yy}$ is general case and is called the anisotropic diffraction~\cite{Polyakov-pre-2002}. Otherwise, that $\delta_{xx}= \delta_{yy}(=1)$ is special case for the ordinary light in the uniaxial crystal or light in the isotropic medium, and the diffraction is isotropic.
By application of an order-of-magnitude analysis method~\cite{Guo-ol-1995}, we have the order of the diffraction terms, respectively 
\begin{equation}\label{order-of-magnitude analysis}
\left|\delta_{xx}\frac{\partial^2\phi}{\partial x^2}\right|\sim\frac{\delta_{xx}}{w_x^2}|\phi|,~~\left|\delta_{yy}\frac{\partial^2\phi}{\partial y^2}\right|\sim\frac{\delta_{yy}}{w_y^2}|\phi|,
\end{equation}
where $w_x$ and $w_y$ are the (dimensionless) beam-width of the optical beam in $x$ and $y$ directions, respectively. Obviously, the beam-expansion will be determined by two factors:  a linear relation with
the coefficients $\delta_{xx}$ and $\delta_{yy}$ that result from the medium anisotropy and an inverse-square relation with the beam-widths that
are determined by
the specific profile of the beam itself propagating in the medium.
We define the radio $\delta_{xx}/\delta_{yy}$ as an anisotropy parameter of the diffraction
%
and assume that $\delta_{xx}>\delta_{yy}$ without loss of generality.

On the propagation for the input beam with a circular shape, the initial beam-expandings are different along the two transverse directions due to the AD, and the beam-expanding in the $x$-direction is stronger than that in the $y$-direction. Thus, the beam will adjust continually its ellipticity during the propagation, and eventually evolves towards an elliptic beam with a specific ellipticity, as discussed for the case along the optical axis of a uniaxial crystal~\cite{Ciattoni-josaa-2002}.
On the other hand, the eigenmodes~\cite{explain}
in the anisotropic media can be directly obtained from the ones in the isotropic medium~\cite{haus-book,Kogelnik-ao-1966} by the scale transform ${x/\delta_{xx}^{1/2}}$ and ${y/\delta_{yy}^{1/2}}$ in Eq.~(\ref{equation-uniaxial-dimensionless}).
The eigenmodes of Eq.~(\ref{equation-uniaxial-dimensionless}) in the Cartesian coordinates are Hermite-Gaussian modes
\begin{eqnarray}\label{Hermite-Gaussian-modes}\nonumber
\phi_{mn}(x,y,z)&=&\frac{C_{mn}}{w}H_m\left(\frac{x}{w_{xhg}}\right)H_n\left(\frac{y}{w_{yhg}}\right)
\exp\left[-\left(\frac{x^2}{2w_{xhg}^2}+\frac{y^2}{2w_{yhg}^2}\right)\right]
\\
&\times&\exp\left[-i\frac{1
}{2R}\left(\frac{x^2}{\delta_{xx}}+\frac{y^2}{\delta_{yy}}\right)+i(m+n+1)\psi\right],
\end{eqnarray}
where $H_m(x)$ is Hermite polynomials, $w_{xhg}=\delta_{xx}^{1/2}w(z)$, $w_{yhg}=\delta_{yy}^{1/2}w(z)$, $w(z)$, $R(z)$, and $\psi(z)$ are given by
$w(z)=\sqrt{1+{z^2}}$, $~R(z)=({1+z^2})/{z}$, and $~\psi(z)=\arctan\left({z}\right)$, respectively,
and $C_{mn}[=(2^{m+n}\pi m!n!\delta_{xx}^{1/2}\delta_{yy}^{1/2})^{-1/2}]$ is the normalization constant which is chosen so that $\int |\phi|^2dxdy=1$. The eigenmodes in the cylindrical coordinates are Laguerre-Gaussian modes, which are like those in the isotropic medium~\cite{Kogelnik-ao-1966} and can be easily written out in the same way above.
It can be observed that the fundamental (basic) eigenmode ($m=n=0)$ in anisotropic medium exhibits transverse asymmetrical structure,
\begin{equation}\label{fundamental mode}
\left|\phi_{00}(x,y,z)\right|=\frac{1}{\sqrt{\pi\delta_{xx}^{1/2}\delta_{yy}^{1/2}}w}\exp\left(-\frac{x^2}{2w_{xhg}^2}-\frac{y^2}{2w_{yhg}^2}\right),
\end{equation}
 as was discussed~\cite{Seshadri-josaa-2003},
while the fundamental mode in the isotropic medium must be transversely circular~\cite{Kogelnik-ao-1966,haus-book}.

\section{Two different propagation cases}
   If the optical beams carry the OAM,
   the elliptically shaped modes even can exist in the media with both linear and nonlinear isotropy~\cite{Desyatnikov-prl-2010,liang-pra-2013}. We have guessed that~\cite{liang-pra-2013} the OAM can result in the effective AD, that is, the OAM enhances the expansion of the beams in the major-axis directions and weakens that in the minor-axis direction. In the following, we will quantitatively
   investigate such an AD induced by the OAM.

   In order to find the quantitative relation between the OAM and its induced AD, we consider the following two cases. Case 1: the propagation
of an elliptic beam without the OAM in the anisotropic media. Case 2: the propagation of an elliptic beam carrying the OAM in the isotropic media.
By comparing the two kinds of propagations, we come to a conclusion about the AD induced by the OAM.

We assume the optical beam at $z=0$ is of the form
\begin{equation}\label{input}
\phi(x,y,0)=\sqrt{\frac{P_{0}}{\pi w_{0x}w_{0y}}}\exp\left(-\frac{x^{2}}{2w_{0x}^{2}}-\frac{y^{2}}{2w_{0y}^{2}}\right)\exp(\textrm{i}\Theta xy),
\end{equation}
where $w_{0x}$ and $w_{0y}$ are two semi-axes,
$P_0=\int\!\!\!\int
|\phi(x,y,z)|^2dxdy$ is the optical power, and $\Theta$ is the cross phase coefficient, which determines the OAM carried by the optical beam~\cite{Desyatnikov-prl-2010,liang-pra-2013}
\begin{equation}\label{OAM}
M=\textrm{Im}\int_{-\infty}^\infty\int_{-\infty}^\infty \phi^{*}\left(x\frac{\partial\phi}{\partial y}-y\frac{\partial\phi}{\partial x}\right)\textrm{d}x\textrm{d}y=\frac{P_{0}}{2}\left(w_{0x}^{2}-w_{0y}^{2}\right)\Theta.
\end{equation}
The solution $\phi(x,y,z)$ of Eq.~(\ref{equation-uniaxial-dimensionless}) with the initial condition~(\ref{input}) can be derived by the
Fresnel diffraction integral~\cite{haus-book} with its anisotropic form
\begin{equation}\label{collins}
\phi(x,y,z)=-i\frac{\exp(iz)}{2\pi z}\int_{-\infty}^\infty\int_{-\infty}^\infty\phi(x',y',0)\exp\left[i\frac{(x-x')^2/\delta_{xx}+(y-y')^2/\delta_{yy}}{2z}\right]dx'dy',
\end{equation}
%
and it is obtained as follows after some complicated calculations
\begin{eqnarray}
\phi(x,y,z)&=&\phi_0\exp\left(-\frac{x^2}{2w_x^2}-\frac{y^2}{2w_y^2}-\frac{xy}{2w_{xy}}+i\varphi\right),\label{solution}
\end{eqnarray}
where
$$\phi_0=\frac{w_{0y}\sqrt{P_0w_{0x}w_{0y}}}{\sqrt{\pi \alpha_r}},~\varphi=c_{xy}xy+c_xx^2+c_yy^2+\frac{\arctan\kappa-\pi}{2},$$
$$w_x=\frac{\delta_{xx}\delta_{yy}}{w_{0x}}\sqrt{\frac{\alpha_r}{w_{0y}^4+\delta_{yy}^2(w_{0x}^2w_{0y}^2\Theta^2+1)z^2}},~
w_y=\frac{\delta_{xx}\delta_{yy}}{w_{0y}}\sqrt{\frac{\alpha_r}{w_{0x}^4+\delta_{xx}^2(w_{0x}^2w_{0y}^2\Theta^2+1)z^2}},$$
$$w_{xy}=-\frac{\alpha_r\delta_{xx}^2\delta_{yy}^2}{2w_{0x}^2w_{0y}^2(w_{0x}^2\delta_{yy}+w_{0y}^2\delta_{xx})\Theta z},~\kappa=\frac{(\delta_{yy}w_{0x}^2+\delta_{xx}w_{0y}^2)z}{w_{0x}^2w_{0y}^2(1-\delta_{xx}\delta_{yy}\Theta^2z^2)-\delta_{xx}\delta_{yy}z^2},$$
\begin{eqnarray}
\nonumber
c_x&=&\frac{z\left[\delta_{xx}\delta_{yy}^2z^2(2w_{0x}^2w_{0y}^2\Theta^2+w_{0x}^4w_{0y}^4\Theta^4)+w_{0y}^4(\delta_{xx}-\delta_{yy}w_{0x}^4\Theta^2)\right]}{2\delta_{xx}^2\delta_{yy}^2\alpha_r},\\ \nonumber
c_y&=&\frac{z\left[\delta_{xx}^2\delta_{yy}z^2(2w_{0x}^2w_{0y}^2\Theta^2+w_{0x}^4w_{0y}^4\Theta^4)+w_{0x}^4(\delta_{yy}-\delta_{xx}w_{0y}^4\Theta^2)\right]}{2\delta_{xx}^2\delta_{yy}^2\alpha_r},\nonumber\\
c_{xy}&=&\frac{w_{0x}^2w_{0y}^2\Theta\left[w_{0x}^2w_{0y}^2-\delta_{xx}\delta_{yy}z^2(w_{0x}^2w_{0y}^2\Theta^2+1)\right]}{\alpha_r\delta_{xx}^2\delta_{yy}^2},\nonumber
\end{eqnarray}
and
$$\alpha_r=\left(2w_{0x}^2w_{0y}^2\Theta^2+1\right)z^4+\left(\frac{w_{0x}^4}{\delta_{xx}^2}+\frac{w_{0y}^4}
{\delta_{yy}^2}\right)z^2+w_{0x}^4w_{0y}^4\left(\frac{1}{\delta_{xx}\delta_{yy}}-\Theta^2z^2\right)^2.$$
\subsection{Case one: elliptic beam without the OAM in anisotropic medium}\label{case one}
\begin{figure}[htb]
\centerline{\includegraphics[width=7cm]{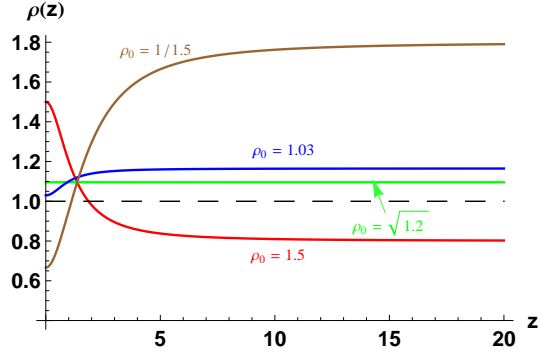}}
\caption{(Color online) Ellipticity functions $\rho(z)$ of elliptic beams propagating in anisotropic media for four different initial $\rho_0$s: $w_{0x}=1.5$ and $w_{0y}=1$, $w_{0x}=\sqrt{1.2}$ and $w_{0y}=1$, $w_{0x}=1.03$ and $ w_{0y}=1$ ($\rho_0>1$), and $w_{0x}=1$ and $w_{0y}=1.5$ ($\rho_0<1$). Parameters are taken to be $\delta_{xx}=1.2,$ and $\delta_{yy}=1$.}\label{evolution-zero-oam}
\end{figure}

To investigate the evolution of the elliptic beam without the OAM in the anisotropic medium, we set $\Theta=0$, then the solution (\ref{solution}) becomes
\begin{equation}\label{elliptic-beam-without-OAM}
\left|\phi(x,y,z)\right|=\phi_0\exp\left(-\frac{x^2}{2w_{xa}^2}-\frac{y^2}{2w_{ya}^2}\right),
\end{equation}
where the two semi-axes are
\begin{equation}\label{wb-1}
w_{xa}^2=w_{0x}^2+\left(\frac{\delta_{xx}}{w_{0x}}\right)^2z^2,~~
w_{ya}^2=w_{0y}^2+\left(\frac{\delta_{yy}}{w_{0y}}\right)^2z^2.
\end{equation}
As mentioned above [see, Eq.~(\ref{order-of-magnitude analysis})], the beam-expansion is determined by both the medium anisotropy and the beam structure. Here, we assume that $\delta_{xx}>\delta_{yy}$ (without loss of generality), and discuss the beam-expanding under two initial conditions that $w_{0x}<w_{0y}$ and $w_{0x}>w_{0y}$, which can be featured by the ellipticity function $\rho(z)$
that is defined as the ratio of the two semi-axes of the elliptical beam and $\rho(z)=w_{xa}(z)/w_{ya}(z)$ here for this case.
Figure \ref{evolution-zero-oam} gives the evolution of $\rho(z)$ for different initial ellipticities $\rho_0[=\rho(0)]$.
In the case that $w_{0x}<w_{0y}$ ($\rho_0<1$), we will have ${\delta_{xx}}/{w_{0x}^2}>{\delta_{yy}}/{w_{0y}^2}$ and $\partial_z\rho(z)>0$. This means that
the stronger initial-beam-expanding in the $x$ direction will make the beam-expanding always be stronger in that direction such that $\rho(z)$ will keep increase. Thus, the elliptic beam will evolve towards a circular beam first,
and again do an elliptic beam but with an ellipticity $\rho(z)>1$  till at a very large distance ($z\rightarrow\infty$),
$\rho(\infty)=\delta_{xx}/\delta_{yy}\rho_0$, as shown by the curve of $\rho_0=1/1.5$ in Fig.\ref{evolution-zero-oam}.
In the case that $w_{0x}>w_{0y}$ ($\rho_0>1$), the elliptic beam will exhibit two different evolutions depending on the relative magnitudes between $\delta_{xx}/{w_{0x}^2}$ and ${\delta_{yy}}/w_{0y}^2$. That $\delta_{xx}/{w_{0x}^2}>{\delta_{yy}}/w_{0y}^2$ ($w_{0x}>w_{0y}$ and $\delta_{xx}>{\delta_{yy}}$) will also yield $\partial_z\rho(z)>0$, the ellipticity $\rho(z)$ will also continuously increase till approach the saturated value $\rho(\infty)$, as can be seen from the curve of $\rho_0=1.03$ in Fig.\ref{evolution-zero-oam}; while if $\delta_{xx}/{w_{0x}^2}<{\delta_{yy}}/w_{0y}^2$, that $\partial_z\rho(z)<0$ can be obtained, thus the beam-expanding will always be weaker in the $x$ direction and $\rho(z)$ will keep decrease, contrary to the case that $w_{0x}<w_{0y}$. The elliptic beam with $\rho>1$ 
will evolve towards a circular beam in the initial phase, and again do an elliptic beam with $\rho<1$,
as shown by the curve of $\rho_0=1.5$ in Fig.\ref{evolution-zero-oam}. Specially, when $\delta_{xx}/w_{0x}^2=\delta_{yy}/w_{0y}^2$, that is
 \begin{equation}\label{case1}
\rho_0=\sqrt{\frac{\delta_{xx}}{\delta_{yy}}},
\end{equation}the total contributions of the beam-expansions due to the medium anisotropy and due to the beam-width are the same in both $x$ and $y$ directions, thus we can have $\partial_z\rho(z)=0$.
In this case, the ellipticity $\rho(z)$ will be constant shown by the curve of $\rho_0=\sqrt{1.2}$ in Fig.\ref{evolution-zero-oam}, and Eq.~(\ref{elliptic-beam-without-OAM}) will be deduced to the fundamental mode, given by Eq.~(\ref{fundamental mode}), of Eq.~(\ref{equation-uniaxial-dimensionless}).
\subsection{Case two: elliptic beam with the OAM in isotropic medium}\label{case two}
The solution for the evolution of the beam carrying the OAM in the isotropic medium can be obtained from Eq.~(\ref{solution}) by setting $\delta_{xx}=\delta_{yy}=1$. Without loss of generality, it is convenient to assume that $w_{0x}>w_{0y}$ in this case. Thus, $\Theta$ can be either positive or negative.
 The cross-sectional shape of the beam will be an oblique ellipse in the $x$-$y$ coordinates for the nonzero OAM. Moreover, the expression of the standard ellipse can be derived  by a coordinate rotation $X=x\cos\theta+y\sin\theta, Y=-x\sin\theta+y\cos\theta, Z=z$, and the rotational angle $\theta$ is determined by
\begin{equation}\label{rotation angle}
\tan(2\theta)=\frac{2w_{0x}^2w_{0y}^2(w_{0x}^2+w_{0y}^2)\Theta z}{(w_{0x}^2-w_{0y}^2)(w_{0x}^2w_{0y}^2-w_{0x}^2w_{0y}^2\Theta^2z^2-z^2)}.
\end{equation}
 Then a major-axis $w_b$ and a minor-axis $w_c$ of the ellipse are found to be
\begin{equation}\label{oneaxis}
\begin{array}{c}
w_b^2=\frac{2w_{xm}^2w_{ym}^2}{w_{xm}^2+w_{ym}^2-\sqrt{(w_{xm}^2-w_{ym}^2)^2+w_{xm}^2w_{ym}^2/w_{xym}^2}}, \\
w_c^2=\frac{2w_{xm}^2w_{ym}^2}{w_{xm}^2+w_{ym}^2+\sqrt{(w_{xm}^2-w_{ym}^2)^2+w_{xm}^2w_{ym}^2/w_{xym}^2}},
\end{array}
\end{equation}
where
$w_{xm}=\left.w_x\right|_{\delta_{xx}=\delta_{yy}=1},w_{ym}=\left.w_y\right|_{\delta_{xx}=\delta_{yy}=1}$ and $w_{xym}=\left.w_{xy}\right|_{\delta_{xx}=\delta_{yy}=1}$, and all of them are functions of the propagation distance $z$.
It can also be observed from Eq.~(\ref{rotation angle}) that the elliptic beam will rotate because of the OAM it carries, the detail of which will be discussed later.

Comparing Eq.~(\ref{oneaxis}) with Eq.~(\ref{wb-1}), we can obtain the effective Fresnel diffraction coefficients
\begin{equation}\label{eff-1}
\delta_{xx}^{\mathrm{eff}}=\frac{w_{0x}\sqrt{w_b^2-w_{0x}^2}}{z},~~
\delta_{yy}^{\mathrm{eff}}=\frac{w_{0y}\sqrt{w_c^2-w_{0y}^2}}{z},
\end{equation}
both of which are generally dependent on the propagation distance $z$. In order to understand what is the physical picture of $\delta_{xx}^{\mathrm{eff}}$ and $\delta_{yy}^{\mathrm{eff}}$, we inspect the incipient effect of the OAM on the beam propagation. For this purpose,
we expand $w_b^2$ and $w_c^2$ in Eq.~(\ref{oneaxis}) with respect to $z$ in a Taylor series about $z=0$ to second order:
\begin{equation}\label{width-expansion}
\begin{array}{c}
w_b^2\approx w_{0x}^2+\frac{1}{w_{0x}^2}\left[1+\frac{w_{0x}^4\left(w_{0x}^2+3w_{0y}^2\right)\Theta^2}
{w_{0x}^2-w_{0y}^2}\right]z^2, \\
w_c^2\approx w_{0y}^2+\frac{1}{w_{0y}^2}\left[1-\frac{w_{0y}^4\left(w_{0y}^2+3w_{0x}^2\right)\Theta^2}
{w_{0x}^2-w_{0y}^2}\right]z^2,
\end{array}
\end{equation}
and their substitution into Eq.~(\ref{eff-1}) yields
\begin{equation}\label{eff-diffraction-expasion}
\delta_{xx}^{\mathrm{eff}}\approx\sqrt{1+\frac{w_{0x}^4\left(w_{0x}^2+3w_{0y}^2\right)\Theta^2}
{w_{0x}^2-w_{0y}^2}},
\delta_{yy}^{\mathrm{eff}}\approx\sqrt{1-\frac{w_{0y}^4\left(w_{0y}^2+3w_{0x}^2\right)\Theta^2}
{w_{0x}^2-w_{0y}^2}}.
\end{equation}
The comparison between the behavior of the elliptic beam with the OAM in the isotropic medium [Eq.(\ref{width-expansion})] and that without the OAM in the anisotropic medium [Eq.~(\ref{wb-1})] tells a story that the former
acts, watched in the $X$-$Y$ coordinate system, as if it propagated in the medium with the Fresnel diffraction coefficients $\delta_{xx}^{\mathrm{eff}}$ and $\delta_{yy}^{\mathrm{eff}}$. Obviously the two coefficients given by Eq.~(\ref{eff-diffraction-expasion}), or more generally Eq.~(\ref{eff-1}), are not equal as long as $\Theta\neq0$. $\delta_{xx}^{\mathrm{eff}}$ and $\delta_{yy}^{\mathrm{eff}}$, therefore, are the anisotropic diffraction induced by the orbital angular momentum.

\begin{figure}[htb]
\centerline{\includegraphics[width=9cm]{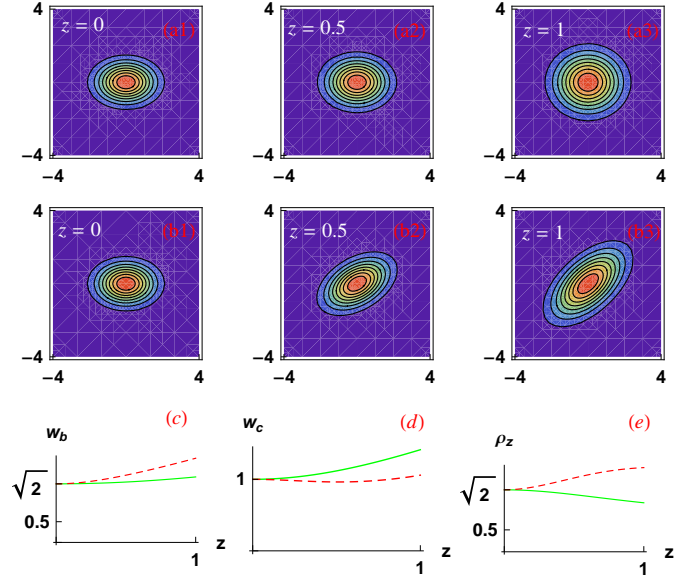}}
\caption{(Color online) Evolutions of the elliptic beams with and without the OAM in the isotropic medium. (a1-a3): zero OAM $\Theta=0$, (b1-b3): with OAM $\Theta=1/2$, (c-d): evolutions of two semi-axes, and (e): evolutions of the ellipticities. $w_{0x}=\sqrt{2}$ and $w_{0y}=1$. Red dashed line: with the OAM, and green solid line: without the OAM. 
}\label{feilinjie-evolution}
\end{figure}

Furthermore, the AD induced by the OAM can be better understood by making a comparison of the propagations of the elliptic beams with and without the OAM in the isotropic medium, as shown in Fig.~\ref{feilinjie-evolution}. For the isotropic medium, the beam-expanding is only inversely proportional to square of the beam width, as discussed in \S\ref{section2}. Thus, the elliptic beam without the OAM spreads more quickly in $y$ (minor-axis) direction than the $x$ (major-axis) direction, resulting in the decrease of its ellipticity;
on the contrary, the ellipticity increases for the elliptic beam with the OAM, as shown in Fig.\ref{feilinjie-evolution}~(e). The reason for such a difference is that the AD induced by the OAM makes the expanding in the $X$ (major-axis) direction be stronger than that in the $x$ direction for the case without OAM and one in the $Y$ (minor-axis) direction be weaker than that in $y$ direction, as shown in Fig.\ref{feilinjie-evolution}~(c-d) and also observed in Eq.~(\ref{eff-diffraction-expasion}).

Setting $\partial_z\rho(z)=0$ where $\rho(z)=w_b(z)/w_c(z)$ that are given by Eq.~(\ref{oneaxis})
yields the critical cross phase coefficient
\begin{equation}\label{case2}
\Theta_c=\pm\frac{w_{0x}^2-w_{0y}^2}{2w_{0x}^2w_{0y}^2}.
\end{equation}
When $\Theta=\Theta_c$, the elliptic beam carrying OAM will evolve with an invariant ellipticity, and the beam profile in the $X$-$Y$ (rotating) coordinate system is the fundamental
mode 
given by Eq.~(\ref{fundamental mode}) with two semi-axes
\begin{equation}\label{cri-width}
(w_b^{cri})^2={w_{0x}^2+\left(\frac{\delta_{xxc}^{\mathrm{eff}}}{w_{0x}}\right)^2z^2},~~
(w_c^{cri})^2={w_{0y}^2+\left(\frac{\delta_{yyc}^{\mathrm{eff}}}{w_{0y}}\right)^2z^2},
\end{equation}
and the AD, $\delta_{xxc}^{\mathrm{eff}}$ and $\delta_{yyc}^{\mathrm{eff}}$, induced by the OAM
\begin{equation}\label{cri-eff-diff}
\delta_{xxc}^{\mathrm{eff}}=\left|\tilde{M}\right|+1+\sqrt{\left|\tilde{M}\right|(\left|\tilde{M}\right|+1)},~~
\delta_{yyc}^{\mathrm{eff}}=\left|\tilde{M}\right|+1-\sqrt{\left|\tilde{M}\right|(\left|\tilde{M}\right|+1)},
\end{equation}
where $\tilde{M}=M_c/P_0$ is the critical OAM per unit power.
Eq.~(\ref{cri-width}) is obtained from Eq.~(\ref{oneaxis}) for $\Theta=\Theta_c$ and Eq.~(\ref{cri-eff-diff}) is from Eq.~(\ref{eff-1}) for $\Theta=\Theta_c$ and a relation [see, Eq.~(\ref{OAM})]
\begin{equation}\label{critical-OAM}
\frac{M_c}{P_{0}}=\frac{1}{2}\left(w_{0x}^{2}-w_{0y}^{2}\right)\Theta_c=\pm\frac{1}{4}\frac{(w_{0x}^2-w_{0y}^2)^2}{w_{0x}^2w_{0y}^2}.
\end{equation}
It can also be found that the two semi-axes given by Eq.~(\ref{cri-width}) satisfy
\begin{equation}\label{ratio-two-axes-rotation}
\frac{w_b^{cri}}{w_c^{cri}}=\left(\frac{\delta_{xxc}^{\mathrm{eff}}}{\delta_{yyc}^{\mathrm{eff}}}\right)^{1/2}
=\sqrt{|\tilde{M}|+1}+\sqrt{|\tilde{M}|}.
\end{equation}The ratio 
$\delta_{xxc}^{\mathrm{eff}}/\delta_{yyc}^{\mathrm{eff}}$ above gives the anisotropic parameter of the AD induced by the OAM.
It tells that an inequation $\delta_{xxc}^{\mathrm{eff}}>\delta_{yyc}^{\mathrm{eff}}$ holds always no matter the sign of the OAM, which means the OAM always enhances the beam-expanding in the major-axis direction and weakens that in the minor-axis direction, and also that the larger the OAM, the stronger the
AD induced by it.

\begin{figure}[htb]
\centerline{\includegraphics[width=7.5cm]{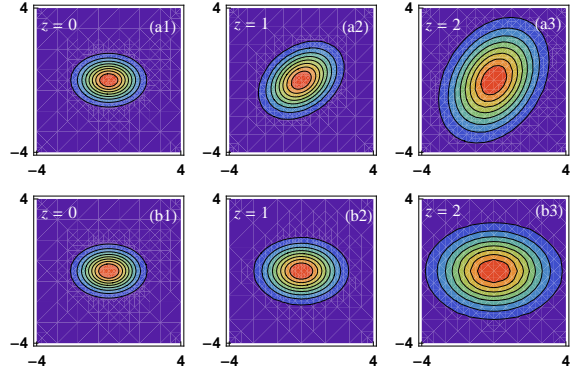}}
\caption{(Color online)  Evolutions of the elliptic beams for two cases. (a1-a3): with the OAM $\tilde{M}=1/8$ in the isotropic medium, (b1-b3): without the OAM in the fictitious anisotropic medium of $\delta_{xx}=3/2$ and $\delta_{yy}=3/4$. The beam widthes are $w_{0x}=\sqrt{2}$ and $w_{0y}=1$.
}\label{evolution}
\end{figure}

Now we can draw a conclusion that observed in the rotating ($X$-$Y$) coordinate system, the elliptic beam in the isotropic medium when its carrying OAM equals the critical value given by Eq.~(\ref{critical-OAM}) behaves exactly same as that with the same widthes in the anisotropic medium with the two coefficients $\delta_{xx}=\delta_{xxc}^{\mathrm{eff}}$ and $\delta_{yy}=\delta_{yyc}^{\mathrm{eff}}$, no matter how long distance they propagate.
Here we give an example to show such a similarity. The input elliptic beam are assumed with $w_{0x}=\sqrt{2}$ and $w_{0y}=1$ and carrying the critical OAM $\tilde{M}=1/8$. Because its carrying OAM induce the AD $\delta_{xxc}^{\mathrm{eff}}=3/2$ and $\delta_{yyc}^{\mathrm{eff}}=3/4$, the elliptic beam can evolve with an invariant ellipticity $\sqrt{2}$ in the isotropic medium, as shown in Fig.\ref{evolution} (a1-a3). 
The fundamental-mode-propagation of the elliptic beam with the same structure but without the OAM in the fictitious anisotropic medium with $\delta_{xx}=3/2$ and $\delta_{yy}=3/4$ also has an invariant ellipticity $\sqrt{2}$, as displayed in Fig.~\ref{evolution} (b1-b3). It can be found the two evolutions are completely equivalent without taking the rotation from the OAM into consideration.

\begin{figure}[htb]
\centerline{\includegraphics[width=10cm]{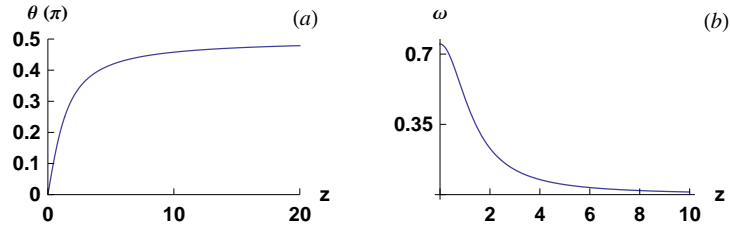}}
\caption{(Color online) Rotation angle $\theta$ (a) and angular velocity $\omega$ (b) of elliptic beams with OAM propagating in isotropic media.
 Parameters are taken to be $w_{0x}=\sqrt{2},w_{0y}=1$ and $\Theta=1/4$. }\label{rotation}
\end{figure}

As mentioned above, the OAM can not only induce the AD, but also make the elliptic beam rotate, as shown in
Figs.~(\ref{feilinjie-evolution}) and (\ref{evolution}).
The rotation angle $\theta$ of the elliptic beam 
can be obtained
 from Eq.~(\ref{rotation angle}) as
\begin{equation}\label{rotation angle2}
\theta=\left\{
    \begin{array}{cc}
 &-\frac{1}{2}\arctan\left[\frac{2 w_{0x}^2 w_{0y}^2 \left(w_{0x}^2+w_{0y}^2\right)\Theta z}{\left(w_{0x}^2-w_{0y}^2\right) \left[z^2+w_{0x}^2 w_{0y}^2 \left(-1+z^2 \Theta ^2\right)\right]}\right],\quad z<\frac{w_{0x}w_{0y}}{\sqrt{w_{0x}^2w_{0y}^2\Theta^2+1}}\\
 &-\frac{1}{2}\arctan\left[\frac{2 w_{0x}^2 w_{0y}^2 \left(w_{0x}^2+w_{0y}^2\right)\Theta z}{\left(w_{0x}^2-w_{0y}^2\right) \left[z^2+w_{0x}^2 w_{0y}^2 \left(-1+z^2 \Theta ^2\right)\right]}\right]+\frac{\pi}{2},\quad z>\frac{w_{0x}w_{0y}}{\sqrt{w_{0x}^2w_{0y}^2\Theta^2+1}}.
    \end{array}
\right.
\end{equation}
The positive OAM ($\Theta>0)$ makes the elliptic beam rotate anticlockwise, otherwise clockwise. When $z\rightarrow+\infty$, rotation angle approaches to $\pm\pi/2$ depending up the sign of the OAM, 
as shown in Fig.~\ref{rotation} (a).
And the angular velocity can be obtained from Eq.~(\ref{rotation angle})
\begin{equation}\label{omega}
\omega\equiv\frac{d\theta}{dz}=\frac{1}{2(1+\tan2\theta)^2}\frac{d}{dz}\left[\frac{w_{xm}^2w_{ym}^2}{w_{xym}(w_{ym}^2-w_{xm}^2)}\right].
\end{equation}
It is known that the OAM is a conservative quantity in isotropic media~\cite{Desyatnikov-prl-2010,liang-pra-2013}. However, because of the diffraction,
the beam width becomes larger as the propagation distance increases, it leads to slowing down of the rotation~\cite{Desyatnikov-prl-2002}, which is shown in Fig.\ref{rotation} (b).
\section{Conclusion}
It is found that the orbital angular momentum carried by the elliptic beam can induce the anisotropic diffraction. By discussing the propagations of the elliptic beam without the OAM in the anisotropic medium and the elliptic beam carrying the OAM in the isotropic medium, we analytically obtain the quantitative relation between the OAM and its induced AD. The OAM always enhances the beam-expanding in the major-axis direction and weakens that in the minor-axis direction no matter the sign of the OAM, and the larger the OAM, the stronger the
AD induced by it. When the elliptic beam without the
OAM propagate in the anisotropic medium, it evolves as the fundamental mode if its ellipticity is the square root of anisotropic parameter $\delta_{xx}/\delta_{yy}$.
And in the isotropic medium, the beam profile in the $X$-$Y$ (rotating) coordinate system is the fundamental
mode if the OAM carried by optical beam equals to its critical value $M_c (\Theta_c)$. The rotation resulting form the OAM is also
discussed, and is found that the absolute value of the rotation angle is no larger than $\pi/2$ and the angular velocity
slows down during the propagation because of the diffraction effect.

Our results give the answer why the elliptic mode can survive in the medium with both linear and nonlinear isotropy where only the mode with cylindrical-symmetry is supposed to exist.
\section*{Acknowledgments}
This research was supported by the National Natural Science Foundation of China, Grant
Nos.~11474109 and 11604199.

\end{document}